\documentclass[groupedaddress,twocolumn,aps,pra,showpacs,fixfloats]{revtex4}
\usepackage{bm}
\usepackage{dcolumn}
\usepackage{graphicx}

\begin{document}

\title{Angle-differential elastic electron scattering off C$_{60}$: A simple
semi-empirical theory versus experiment}

\author{M. Ya. Amusia$^{1, 2}$}
\author{L. V. Chernysheva$^{2}$}
\author{ V. K. Dolmatov$^{3}$}

\affiliation{$^1$Racah Institute of Physics, Hebrew University, 91904 Jerusalem, Israel}
\affiliation{$^2$A. F. Ioffe Physical-Technical Institute, 194021 St. Petersburg, Russia }
\affiliation{$^3$Department of Physics and Earth Science, University of North Alabama, Florence, AL 35632, USA }

\begin{abstract}
We prove in the present paper that a simple modeling of a complicated
highly polarizable C$_{60}$ target by a rectangular (in a radial coordinate)
square well potential in combination with the static polarization potential
provides a viable approximation for a low-energy elastic electron scattering
off this target. The proof is based on the results of the comparison of the
calculated angle-differential elastic electron scattering cross section off
C$_{60}$ versus corresponding experimental data.
\end{abstract}

\maketitle

Low-energy elastic electron scattering off a C$_{60}$ fullerene
or $A$@C$_{60}$ endohedral fullerene has been in the focus of
theorists for some time now, see, e.g.,
\cite{Gianturco,Winstead,e+C60(1),e+C60(3),e+C60(2),e+C60(A),e+C60(4),Msezane}
and references therein. Whereas works~\cite{Gianturco,Winstead}
addressed the problem from the first ab initio principles, a
part of the work \cite{Winstead} as well as works
\cite{e+C60(1),e+C60(2),e+C60(3),e+C60(4)} used a semi-empirical
approach where the C$_{60}$ cage was modeled by a rectangular
(in a radial coordinate) square well potential, $U_{\rm C}(r)$,
of certain inner radius, $r_{0}$, thickness, $\Delta $, and
depth, $U_{0}$, to address a low-energy elastic $e^{-} + {\rm
C_{60}}$ scattering \cite{Winstead,e+C60(1),e+C60(2),e+C60(3)}
or $e^{-} + A@{\rm C}_{60}$ scattering \cite{e+C60(3),e+C60(4)}:
\begin{eqnarray}
U_{\rm C}(r)=\left\{\matrix {
-U_{0}, & \mbox{if $r_{0} \le r \le r_{0}+\Delta$} \nonumber \\
0 & \mbox{otherwise.} } \right.
\label{SWP}
\end{eqnarray}
Works \cite{Winstead,e+C60(1),e+C60(2)} suffered a serious
drawback because of not accounting for polarization of a highly
polarizable C$_{60}$ cage by a scattering electron; the dipole
static polarizability, $\alpha_{\rm C}$, of C$_{60}$ is
approximately $850$ $a.u.$ \cite{Amusia_alpha} ($a.u.$ stands
for ``atomic units'' which is the system of units used
throughout the present paper unless stated otherwise). Further
corrections to the model, made in works
\cite{e+C60(3),e+C60(4)}, improved the approximation
considerably by introducing a new model potential, $U_{\rm C
\alpha} (r)$, of a polarizable C$_{60}$ as the sum of the
$U_{C}(r)$ potential and the static polarization potential
$V_{\alpha}(r)$:
\begin{eqnarray}
U_{\rm C\alpha}(r) =  U_{\rm C}(r) + V_{\alpha}(r),
\label{UCalpha}
\end{eqnarray}
\begin{eqnarray}
V_{\alpha}(r)=-\frac{\alpha_{\rm C}}{2(r^2 + b^2)^2}.
\label{Valpha}
\end{eqnarray}
Here, $b$ is a cut-off parameter of the order of the size of the
target (C$_{60}$, in our case) to prevent divergence of the
potential at $r\rightarrow 0$. In the present paper, the $b$
parameter is put to $b\approx 8$ which is approximately the size
of C$_{60}$.

The polarization potential $V_{\alpha }(r)$ originates from
physics of atom, where it was introduced in studies of a
concrete negative ion O$^{-}$ with $\alpha_{\rm C}$ treated as
an adjustable parameter in \cite{Bates} yet in $1943$. This
potential was developed by introducing energy dependence of
$\alpha_{\rm C}$, but has remained a phenomenological one
\cite{Drukarev}. A new approach, based on a many-body
diagrammatic technique suitable for atoms, leads to good results
without any adjustable parameters (see \cite{ACYa} and
references therein).

Although quite accurate, the \textit{ab initio} approach is
difficult to apply to such a tremendously complex objects as
fullerenes or endohedrals. This justifies using an approach that
is much less accurate but at the same time much simpler
\cite{e+C60(3)}. However, in the absence of the comparison with
corresponding experimental data for elastic $e^{-} + \rm C_{60}$
or $e^{-} + A@\rm C_{60}$ scattering it has remained unclear how
viable are the approximations above, Eqs.~(\ref{SWP}) --
(\ref{Valpha}), in the application to the scattering process in
question. Whereas we are not aware of the existence of
experimental data for the integral, i.e, total elastic
scattering cross section, $\sigma_{\rm tot}$, we have become
aware of a long-time existing experimental data \cite{Tanaka}
for the angle-differential elastic scattering cross section,
$\frac{d\sigma }{d\Omega }$, in the energy region up to about
$12$ eV of the electron impact energy. It is the aim of the
present paper to demonstrate the viability of the simple
phenomenological model,
Eqs.~(\ref{SWP})--(\ref{Valpha}). To meet the goal, we calculate
in the present work $\frac{d\sigma }{d\Omega }$ for $e^{-} + \rm
C_{60}$ scattering, obtained with the utilization of
Eqs.~(\ref{UCalpha}) and (\ref{Valpha}), and compare the thus
obtained result with experiment \cite{Tanaka} for a number of
scattering angles $\theta =30,70,80$ and $90^{\mathrm{o}}$. We
demonstrate a reasonably good agreement between measured and
calculated data.

The angle-differential scattering cross section is calculated
using the well-known formula, see, e.g., \cite{LL}:
\begin{eqnarray}
\frac{d\sigma }{d\Omega } &=&\frac{1}{k^{2}}\sum_{\ell ,\ell ^{\prime
}=0}^{\infty }(2\ell +1)(2\ell ^{\prime }+1)\sin \delta _{\ell }\sin \delta
_{\ell ^{\prime }}  \nonumber \\
&&\times \cos (\delta _{\ell }-\delta _{\ell ^{\prime }})P_{\ell }(\cos
\theta )P_{\ell ^{\prime }}(\cos \theta ).  \label{dsigma}
\end{eqnarray}
Here, and everywhere else in the paper, $k$ is the electron
momentum, $\theta $ and $\Omega $ are the scattering angle and
solid angle, respectively, $\delta_{\ell }$s are scattering
phase shifts and $P_{\ell}(\cos \theta )$ is the Legendre
polynomial of the $\ell $th order.

The total electron elastic-scattering cross section, $\sigma_{\rm tot}(\epsilon)$, is calculated in accordance with the standard formula for
electron scattering by a central-potential field:
\begin{eqnarray}
\sigma_{\rm tot}(k)= \frac{4\pi}{k^2}\sum^{\infty}_{\ell=0}(2\ell+1)\sin^{2}\delta_{\ell}(k).
\label{sigma}
\end{eqnarray}

\begin{figure}[tbp]
\includegraphics[width=8cm]{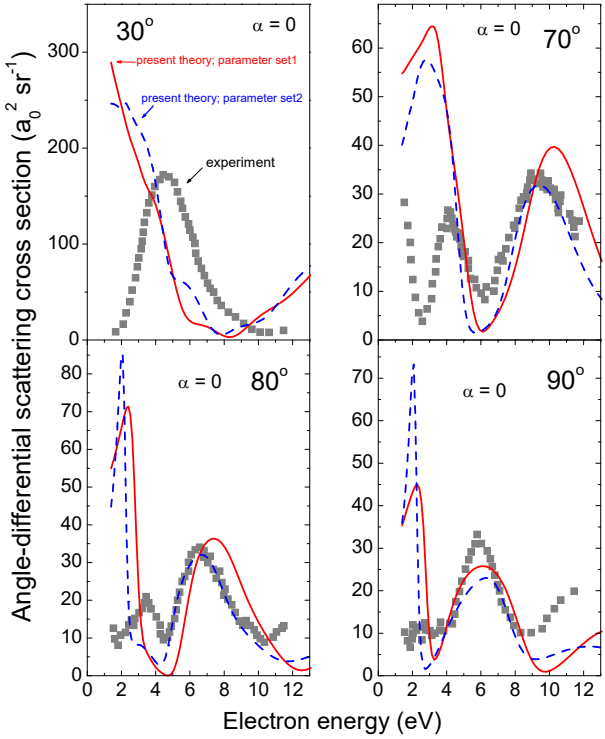}
\caption{(Color online) $e^{-} + \rm C_{60}$
$\frac{d\protect\sigma }{d\Omega }$ (in units
$a_{0}^{2}sr^{-1}$, $a_{0}$ being the first Bohr radius)  at
$\theta =30$, $70$, $80$ and $90^{\rm o}$. Solid and dashed
lines, results of the present calculation obtained
\textit{without} account for polarizability of C$_{60}$ with the
use of, respectively, the set$1$ and set$2$ for $r_{0}$, $\Delta
$ and $U_{0}$. Solid squares, experiment \cite{Tanaka}. Note,
the experimental points are copied from \cite{Gianturco} ``as
is'' in order to not distort the comparison of theory
\cite{Gianturco} with experiment in other parts of the present
paper.} \label{DCS0}
\end{figure}
\begin{figure}[tbp]
\includegraphics[width=8cm]{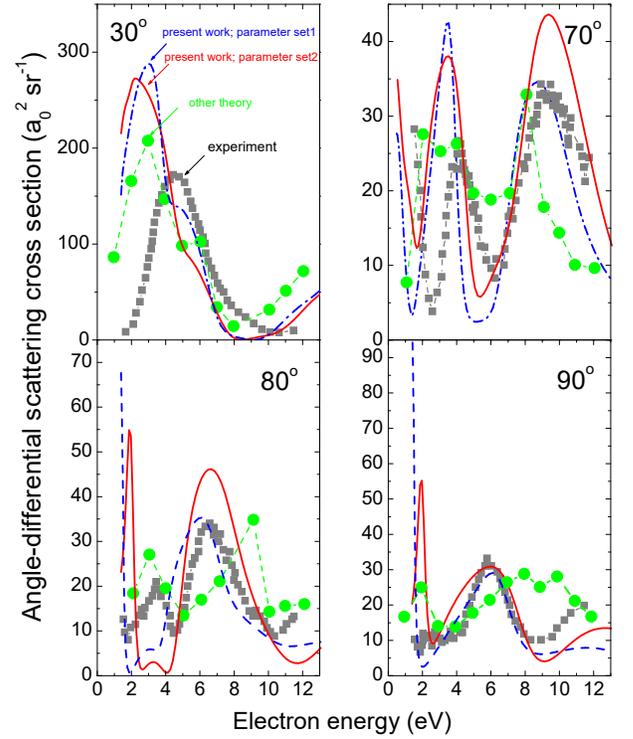}
\caption{(Color online) $e^{-} + \rm C_{60}$ $\frac{d\sigma
}{d\Omega }$ (in units $a_{0}^{2}sr^{-1}$, $a_{0}$ being the
first Bohr radius) at $\theta =30$, $70$, $80$ and
$90^{\mathrm{o}}$. Solid and dashed lines, results of the
present calculation obtained \textit{with} account for
polarization potential of C$_{60}$ with the use of,
respectively, the set$1$ and set$2$ for $r_{0}$, $\Delta $ and
$U_{0}$. Solid squares, experiment \cite{Tanaka}. Solid circles,
results of the \textit{ab initio} theory \cite{Gianturco}. As in
Fig.~\ref{DCS0}, both the experimental points (solid squares)
and results of the \textit{ab initio} theory \cite{Gianturco}
(solid circles) have been copied from \cite{Gianturco} ``as is''
in order to not distort the comparison of theory
\cite{Gianturco} with experiment in any way.} \label{DCS}
\end{figure}
The radial parts, $R_{\epsilon \ell}(r)$, of the wave functions
of scattering states with the definite energy $\epsilon =
k^{2}/2$, $\Psi_{\epsilon,\ell,m_{\ell}}(r) =
(R_{\epsilon\ell}(r)/r)Y_{\ell,m_{\ell}}(\theta,\phi)$, are
obtained by solving the corresponding radial Schr\"{o}dinger
equation:
\begin{eqnarray}
-\frac{1}{2}\frac{d^2R_{\epsilon \ell}}{dr^2} +\left [\frac{\ell(\ell+1)}{2
r^2} + U_{\rm C\alpha}(r) \right ]R_{\epsilon\ell}(r)  \nonumber \\
= E_{\epsilon \ell} R_{\epsilon\ell}(r).
\label{EqShr}
\end{eqnarray}
Once $R_{\epsilon \ell}(r)$ are determined, the needed electron
elastic-scattering phase shifts, $\delta_{\ell}(k)$, are found
by referring to $R_{k\ell}(r)$ at $r \gg 1$:
\begin{eqnarray}
R_{k\ell}(r) \rightarrow \sqrt{\frac{2}{\pi}}\sin\left(k r -\frac{\pi\ell}{2}
+\delta_{\ell}(k)\right).  \label{P(r)}
\end{eqnarray}
The determination of the scattering phase $\delta_{\ell }(k)$
from a numerically derived wave function $R_{k\ell }(r)$
requires a special programm, presented in \cite{AmCher}. In the
calculations, we use two different sets of the adjustable
parameters $r_{0}$, $\Delta $ and $U_{0}$ for the $U_{\rm C}$
cage potential, Eq.~(\ref{SWP}). One of the two sets of the
parameters, referred to as the ``set$1$'', is $r_{0}^{(1)}=5.8$,
$\Delta ^{(1)}=1.9$, $U_{0}^{(1)}=0.302$ $a.u.$, as in, e.g.,
Refs.~\cite{JPCVKD} and references therein, the other one,
referred to as the ``set$2$'', is $r_{0}^{(2)}=5.26$, $\Delta
^{(2)}=2.91$, $U_{0}^{(2)}=0.260$ $a.u.$, as in, e.g.,
Refs.~\cite{e+C60(4),Winstead} and references therein. The use
of two sets of parameters permits to test the sensitivity of the
results to parameters variation. The utilization of the energy
independent $U_{\rm C\alpha}(r)$, as well as the omission of
electron exchange potential, simplifies the calculations
drastically. Contrary to the \textit{ab initio} many-body
approach \cite{AmCherep}, potential $U_{\rm C\alpha}(r)$ permits
easily calculate scattering phases $\delta_{\ell }$ almost
irrespectively to the value of $l$. So, we performed a trial
calculation that showed that $32$ partial electronic waves with
the orbital momentum $\ell $ up to $\ell =31$ is the enough
number of terms to be accounted in the calculation of
$\frac{d\sigma }{d\Omega }$ in order to ensure the convergence
of the sum in Eqs.~(\ref{dsigma}) and (\ref{sigma}).

Corresponding calculated data for $\frac{d\sigma }{d\Omega }$,
obtained without account for the $V_{\alpha }$ polarization
potential potential, Eq.~)\ref{Valpha}), i.e., with the use of
$U_{\rm C}(r)$ determined by Eq.~(\ref{SWP}) alone, are depicted
in Fig.~\ref{DCS0}, whereas those obtained with the account for
the $V_{\alpha }$ potential, i.e., with the use of $U_{\rm
C\alpha}(r)$, Eq.~(\ref{UCalpha}), are depicted in
Fig.~\ref{DCS}.
\begin{figure}[tbp]
\includegraphics[width=8cm]{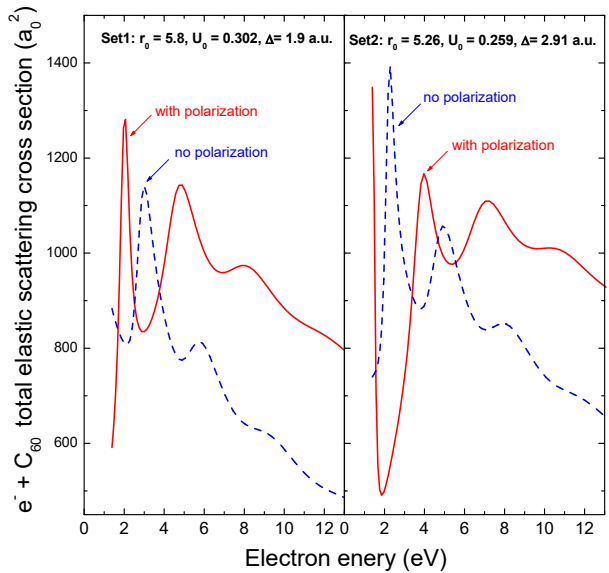}
\caption{(Color online)  $e^{-} + \rm C_{60}$ total elastic
scattering cross section (in units $a_{0}^{2}$, $a_{0}$ being
the first Bohr radius) calculated with and without account for
polarizability of C$_{60}$. Solid line (dashed line) - results
obtained with the use of the parameter-set$1$ (parameter-set$2$)
in the calculation, respectively.} \label{sigmas}
\end{figure}
One can see, that the account for C$_{60}$ polarization
potential, Fig.~\ref{DCS}, corrects $\frac{d\sigma }{d\Omega }$
considerably, especially at $\theta =30$ and $70^{\mathrm{o}}$,
making the agreement between the present theory and experiment
be quite reasonable. This agreement, the authors dare to state,
is, in the whole, even somewhat better than that between the
\textit{ab initio} theory \cite{Gianturco} and experiment. As
for the comparison of results, obtained with the use of the
parameter-set$1$ and parameter-set$2$  for $r_{0}$, $U_{0}$, and
$\Delta $), it is difficult to give an unambiguous preference to
the one over the other.

To let the reader get a better insight into differences between
a static (``frozen'', $\alpha =0$) and polarizable ($\alpha \neq
0$) C$_{60}$ fullerene cage, in Fig.~\ref{sigmas} depicted are
the $e^{-} + \rm C_{60}$ elastic total scattering cross sections
calculated with and without account for polarizability of
C$_{60}$ in the approximations described above.
One can see that the effect of C$_{60}$ polarization potential
acting upon the elastically scattered electron alters the cross
section considerably.

The usual estimation of the required upper value of the partial wave taken
into account is $l_{\max }\simeq kr_{0}$. In the considered case,
$l_{\max}\simeq 6$  even for the highest considered incoming electron energy.
This is well below the upper $l$ value 32, included in our calculations. This
estimation is valid for all short-range potentials, namely those that
decrease faster than $~1/r^{2}$ with distance $r$ growth. Note, however, that
the $d\sigma/d\Omega$  requires more phases than the total cross-section
$\sigma$, since the smallest phase $\delta_{l_{max}}$ that is taken into
account contributes to $d\sigma/d\Omega$ a term $~\sin\delta_{{l}_{max}}$ ,
while to $\sigma$ the contribution of this same phase is much smaller, since
it is $~(\sin\delta_{l_{max}})^{2}\ll\sin\delta_{l_{max}}$ for
$\delta_{l_{max}}\ll1$.

In conclusion, the provided in the present paper results prove the
applicability of the $U_{\rm C\alpha}$ potential (\ref{UCalpha}) to
tackling problems of electron scattering off a highly polarizable C$_{60}$
fullerene. Note that, obviously, by a fine-tuning of the parameters $r_{0}$,
$U_{0}$ and $\Delta $, as well the polarization parameters $\alpha $ and $b$,
one can achieve a yet better agreement between theory and experiment than
the present one. One would have been too naive to have expected more from
the utilized simple modeling of C$_{60}$, and yet its surprising viability
is obvious, at least in the aim of getting the initial insight into the
polarization potential effects in electron scattering of C$_{60}$.

\end{document}